\documentclass[twocolumn,prb,aps]{revtex4}
\usepackage{graphicx}
\usepackage{dcolumn}
\usepackage{amsmath}
\newcommand{\figurewidth}{8.3cm}

\begin{document}
%\scriptsize
%% LaTeX will automatically break titles if they run longer than
%% one line. However, you may use \\ to force a line break if
%% you desire.

\title{{\scriptsize \sf \vspace*{-23pt}Manuscript submitted to Astrophys. J. Lett. (2008).\\[10pt]}
A Massive Core in Jupiter Predicted From First-Principles Simulations}

%% Use \author, \affil, and the \and command to format
%% author and affiliation information.
%% Note that \email has replaced the old \authoremail command
%% from AASTeX v4.0. You can use \email to mark an email address
%% anywhere in the paper, not just in the front matter.
%% As in the title, use \\ to force line breaks.

\author{B. Militzer}
\affiliation{Departments of Earth and Planetary Science and of Astronomy, University of California, Berkeley, CA 94720, USA.}
%\email{militzer@berkeley.edu}

\author{W. B. Hubbard}
\affiliation{Lunar and Planetary Laboratory, The University of Arizona, Tucson, AZ 85721, USA.}
%\email{hubbard@lpl.arizona.edu}

\author{J. Vorberger}
\affiliation{Department of Physics, University of Warwick, Coventry CV4 7AL, UK.}

\author{I. Tamblyn and S.A. Bonev}
\affiliation{Department of Physics and Atmospheric Science, Dalhousie University, Halifax NS B3H 3J5, Canada.}

\begin{abstract}
Hydrogen-helium mixtures at conditions of Jupiter's interior are
studied with first-principles computer simulations. The resulting
equation of state (EOS) implies that Jupiter possesses a central core
of 14 -- 18 Earth masses of heavier elements, a result that supports
core accretion as standard model for the formation of hydrogen-rich
giant planets. Our nominal model has about 2 Earth masses of planetary
ices in the H-He-rich mantle, a result that is, within modeling
errors, consistent with abundances measured by the 1995 Galileo Entry
Probe mission (equivalent to about 5 Earth masses of planetary ices
when extrapolated to the mantle), suggesting that the composition
found by the probe may be representative of the entire
planet. Interior models derived from this first-principles EOS do not
give a match to Jupiter's gravity moment $J_4$ unless one invokes
interior differential rotation, implying that jovian interior dynamics
has an observable effect on the measured gravity field.
\end{abstract}


\maketitle

\section{Introduction}

The discovery of over two hundred extrasolar planets resembling the
giant planet Jupiter in mass\footnote{Schneider, J., the Extrasolar Planets Encyclopaedia http://exoplanet.eu} and
composition~\citep{citation2} has raised fundamental questions about
the origin and inner structure of these bodies.   Establishing the existence
of a dense core in Jupiter is vital for understanding the key
processes of planetary formation. An observed
correlation between the metallicity of the parent star and
the likelihood of giant planets orbiting it~\citep{citation3}
may indicate that an initial core aggregated from solid
planetesimals triggers the gravitational collapse of nebular gases to
form giant planets~\citep{citation4}, or alternatively, increased metallicity enhances
the direct collapse of giant planets from the nebula~\citep{citationX}.
Current planetary
models~\citep{SC95,citation6} predict only a very small
core between 0 and 7 Earth masses ($M_{\rm E}$) for Jupiter, lending
support to core-accretion theories with comparatively small cores~\citep{Po96},
late-stage core erosion scenarios~\citep{Gu04}, or
suggesting that jovian planets are able to form directly from gases
without a triggering core~\citep{citation7}. Using first-principles
simulations for hydrogen-helium mixtures at high pressure, we show
here that Jupiter possesses a significant central core of 16$\pm$2
$M_{\rm E}$ of heavier elements, a result that supports core accretion
as the standard model for formation of hydrogen-rich giant planets.
In our model, Jupiter's mantle, defined as the H-He-rich outer
layers, shows no enrichment of heavier
elements, indicating that compositional measurements from the 1995
Galileo Entry Probe mission may be representative of most of the
planet. The derived interior models also provide the first evidence
that Jupiter's interior does not rotate as a solid body because no
match to Jupiter's gravity moment $J_4$ is obtained unless one invokes
interior differential rotation, implying that jovian interior dynamics
have an observable effect on the measured gravity field.

While laboratory techniques cannot yet probe deep into Jupiter's
interior (reaching pressures P $\sim$ 100--1000 GPa and temperatures T
$\sim$ 10000 K), advances in first-principles computer simulation
techniques have made it possible to characterize the equation of state
(EOS) of hydrogen-helium mixtures for the entire planet. The
determination of the EOS, in combination with the modeling reported
here, enables us to more precisely specify Jupiter's mantle
metallicity and core mass.

\section{Simulation Methods}

All density functional molecular dynamics (DFT-MD) simulations were
performed with either the CPMD code~(\cite{CPMD}) using local
Troullier-Martins norm-conserving pseudopotentials or with the Vienna
{\it ab initio} simulation package~\citep{VASP1} using the projector
augmented-wave method~\citep{PAW}. The nuclei were propagated using
Born-Oppenheimer molecular dynamics with forces derived from the
electronic ground state. Simulations with thermally excited electronic
states did not yield any significant correction to the EOS in
Jupiter's interior. Exchange-correlation effects were described by the
generalized gradient approximation~\citep{PBE}. The electronic
wavefunctions were expanded in a plane-wave basis with a 35-50 Hartree
energy cut-off. The simulations were run for 2 picoseconds with time
steps ranging between 0.2 and 0.8 femtoseconds. We performed well over
a hundred separate DFT-MD simulations on a non-uniform grid in density
and temperature ranging from $\rho$=220 -- 6006 kgm$^{-3}$ and 500 --
20000 K. At lower densities, we used classical Monte Carlo simulations
with fitted potentials~\citep{citation30}, which are in good very
agreement with the SC EOS.

All simulations were performed with 110 hydrogen and 9 helium atoms in
periodic boundary conditions. The initial simulations were performed
with $\Gamma$ point sampling of the Brillouin zone only. Further
simulations with uniform grids of up to 4x4x4 k-points exhibited a
correction to the pressure in the metallic regime. This correction
slowly increases with density, eventually reaching $-1.6$\% at
conditions of Jupiter's core. Finite size effects were further
analyzed by~\citet{Vo07}. The isentropes were derived from a
thermodynamically consistent fit to the free energy using an approach
that was described in detail in Ref.~\citep{Mi08}.

\begin{figure}[!]
%\epsscale{.80}
\includegraphics[angle=0,width=\figurewidth]{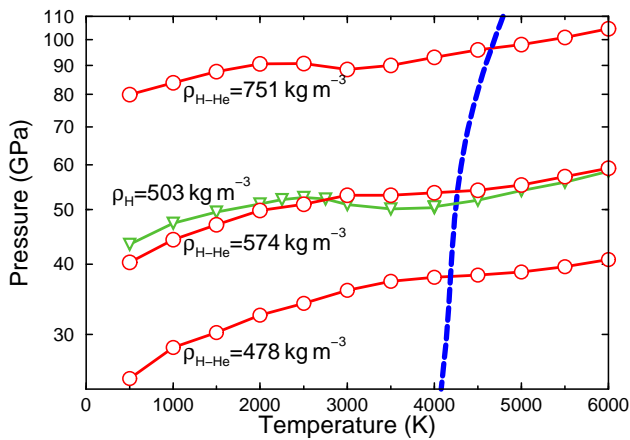}
\caption{Isochores derived from DFT-MD simulations of
       H-He mixtures (circles, Y=0.2466) and pure hydrogen
       (triangles). Results for mixtures predict a positive
       Gr\"uneisen parameter, $(\partial P/\partial T)|_V > 0$, along
       Jupiter's isentrope (dashed line) predicting the planet to
       be isentropic and fully convective.
\label{P_vs_T}}	
\end{figure}

%%% fig 1

DFT-MD predicts a continuous transition from the molecular to a
dissociated regime in fluid hydrogen as function of
pressure~\citep{Vo07,Vo07b}, which is consistent with quantum Monte
Carlo simulations~\citep{delaney06}. The dissociation transition in
dense hydrogen is driven by the increasing overlap of molecular
orbitals. At sufficiently high compression, Pauli exclusion effects
trigger a delocalization of the electronic charge~\citep{citation31},
the band gap closes~\citep{Vo07,Vo07b}, the conductivity increases,
the protons are no longer paired, and eventually the system assumes a
metallic state~\citep{citation15}. As the temperature of the fluid
increases, this transition occurs at lower pressures due to stronger
collisions. When helium is added to dense hydrogen, it localizes the
electronic charge because of its stronger binding.  It dilutes the
hydrogen subsystem by reducing overlap of molecular orbitals,
increasing the stability of hydrogen molecules compared to pure
hydrogen at the same $P$ and $T$~\citep{Vo07}.

While the DFT-MD EOS has a positive compressibility for all $P$ and
$T$, such that $(\partial P/\partial \rho)_T > 0$ (where $\rho$ is the
mass density), the method predicts a negative Gr\"uneisen parameter,
equivalent to $(\partial P/\partial T)_\rho < 0$, for the dissociation
region of pure hydrogen~\citep{Vo07,Vo07b}.  A negative Gr\"uneisen
parameter would introduce a convection barrier into Jupiter's
mantle. However, Figure~\ref{P_vs_T} demonstrates that in a H-He
mixture, the region where $(\partial P/\partial T)_\rho$ is negative
occurs at higher pressures and significantly lower temperatures than
are present in Jupiter's interior.  Consequently, Jupiter's mantle is
predicted to be fully convective and isentropic.

\section{Interior Models}

Were Jupiter of exactly solar composition and all its
magnesium-silicates and iron together with the abundant hydrides
CH$_4$, NH$_3$, and H$_2$O concentrated in a dense core, considering
recent reductions in the solar C and O abundances, its core mass would
only comprise about 3 $M_{\rm
E}$~\citep{citation8,citation9,citation10}. The core mass thought to
trigger nebular collapse is several times
larger~\citep{citation4}. Detection of a central core with a
resolution of a few $M_{\rm E}$, only~1\% of Jupiter's total mass of
318 $M_{\rm E}$, places stringent demands on the accuracy of the
hydrogen-helium EOS. We show in this Letter that an EOS based on
density functional molecular dynamics (DFT-MD) differs from the
previous Saumon-Chabrier-Van Horn (SC) EOS~\citep{SC95,citation11} by
considerably more than one percent.

\begin{figure}[!]
\includegraphics[angle=0,width=\figurewidth]{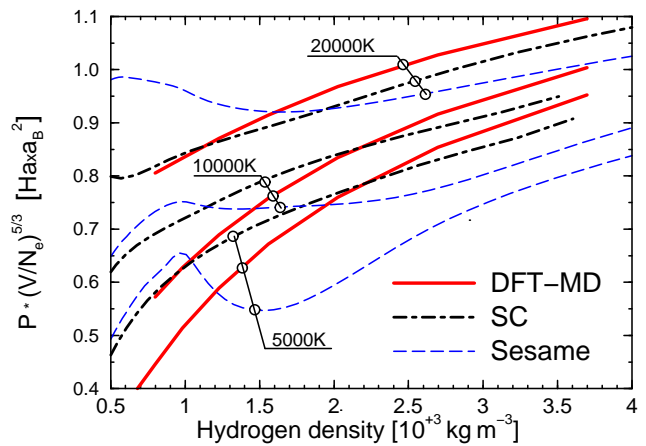}
\caption{Comparison of the DFT-MD EOS with the SC and Sesame models. Three isotherms
       for pure hydrogen are shown in the metallic regime at high pressure. $P$ is scaled by the volume per electron to
       the power $\frac{5}{3}$ to remove most of the density
       dependence.
\label{fig1a}}	
\end{figure}

%fig 2

Figure~\ref{fig1a} compares the DFT-MD results for pure H with the SC
EOS. The SC model relies on analytical techniques that describe
hydrogen as an ensemble of stable molecules, atoms, free electrons,
and protons. Approximations are made to characterize their
interactions. Conversely, with DFT-MD one simulates a fully
interacting quantum system of over a hundred electrons and nuclei. The
isotherms shown in Fig.~\ref{fig1a} specify the EOS in the inner
regions of Jupiter where hydrogen is metallic. Due to the lack of
experimental data at this density, predictions from chemical models
including the Sesame database~\citep{Sesame} vary substantially. The
DFT-MD method obtains higher densities than the SC EOS for much of the
pressure range characteristic of the interior of Jupiter. This has
great influence on the predicted core mass. Moreover, DFT-MD predicts
a continuous molecular-to-metallic transition while the original SC
EOS model predicts this transition to be of first order, which
introduced the possibility of having different chemical compositions
in the inner and outer layers of Jupiter. \citet{citation17} showed
that having flexibility to redistribute materials allows one to make a
coreless model for Jupiter while typical models with constant chemical
composition in the mantle predict a core of about 7 $M_{\rm E}$ when
the SC EOS is used.  Conversely, our DFT-MD EOS predicts Jupiter's
mantle to be isentropic, fully convective, and of constant chemical
composition.  As we show below, the mantle metallicity is so low that
compositional gradients would be unlikely to change this conclusion.

\begin{figure}[t]
\includegraphics[angle=0,width=\figurewidth]{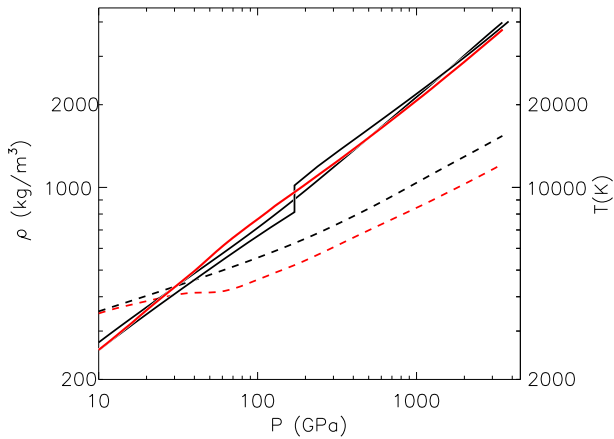}
\caption{Comparison of the resulting Jupiter
       models. Solid curves show mass density (left-hand ordinate) and
       dashed curves show temperature T (right-hand ordinate), as a
       function of pressure $P$. In contrast to Fig.~\ref{fig1a}, the
       DFT-MD (red line) and the SC models have differing chemical
       compositions and temperatures at each pressure. The outermost
       layers at $P<$ 100 GPa make about 90 percent of the
       contribution to the gravity moment $J_4$. The black curves show
       two models based on the smoothed and discontinuous versions of
       the SC EOS.
\label{fig1b}}
\end{figure}

Jupiter isentropes from different models are compared in
Fig.~\ref{fig1b}. The chemical composition corresponding to each
isentrope was determined by adding $\sim 0.7$ mass percent of CH$_4$, NH$_3$,
and H$_2$O (collectively called ``planetary
ices''~\citep{citation18}), using ideal mixing, to the mixture of H
and He considered here.  The adequacy of the ideal-mixing
approximation was verified by several H-He simulations at relevant
pressures and temperatures with ``guest atoms'' of C, N or O.  The
DFT-MD EOS was derived with a single H-He composition, corresponding
to 9 He atoms per 110 H atoms, or a He mass fraction $Y$ = 0.2466, to
be contrasted with the Jovian atmospheric He mass fraction measured by
the Galileo entry probe, $Y$ = 0.2315 $\pm$
0.0061~\citep{citationY}. 
The DFT-MD isentrope was then perturbed to
the probe He abundance by using ideal mixing with a pure He density
at the same $T$ and $P$, synthesized from separate DFT-MD simulations
for pure helium~\citep{Mi08}. As can be seen in Fig.~\ref{fig1b},
DFT-MD implies lower temperatures in the deep jovian interior in
isentropic models, further contributing to generally increased density
with respect to SC models. The high density requires us to adopt $Y$ = 0.2315
throughout the mantle in order to allow a nonnegative planetary ice fraction.
Figure~\ref{fig1b} compares the original
discontinuous SC EOS as well as the smoothed SC EOS version that we
use in Figs.~\ref{fig1a} and discuss below.
 
The DFT-MD EOS yields a significant revision of the interior structure
of Jupiter. The resulting model predicts a large central core of 16
Earth masses and a low metallicity for the mantle while models based
on the SC EOS imply a small core and a much higher metallicity in the
mantle, comprising some tens of Earth masses of elements heavier than
H and He~\citep{citation6,citation17}. DFT-MD based models predict
that the planetary ices were primarily incorporated into a massive
solid core and depleted gaseous nebula. Conversely, models based on
the SC EOS imply that planetary ices were largely accumulated along
with the nebular hydrogen and helium when Jupiter
formed. Figure~\ref{fig1b} shows that the Jupiter model based on the
DFT-MD EOS is about 5\% denser than the model based on the smoothed SC
EOS over the pressure range 30 GPa $<$ $P$ $<$ 300 GPa, which spans
about 30\% of the mass of Jupiter, while it is a few percent less
dense at deeper layers. The Galileo Probe found abundances of CH$_4$,
NH$_3$, and H$_2$O (at the deepest level) corresponding to
at most about 5 $M_{\rm E}$ of
planetary ices if extrapolated to the entire mantle.  Our nominal model
has about 2 $M_{\rm E}$ of planetary ices in the mantle.  Given the cumulative
uncertainties of the DFT-MD EOS and the planetary ices EOS, the
nominal model is consistent with such an extrapolation of the Galileo
measurements.

\begin{figure}[t]
\includegraphics[angle=0,width=\figurewidth]{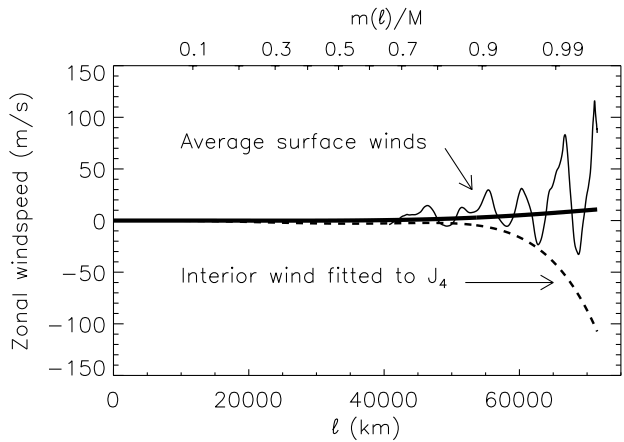}
\caption{Jupiter's zonal wind speeds as a function of $l$,
       the distance from the rotation axis. The oscillating curve
       shows an average of northern and southern hemisphere
       surface zonal winds as determined by cloud motions; the heavy
       solid curve shows a fit of the surface winds with an
       eighth-order polynomial in $l$. The dashed curve shows our
       preferred rotation model that provides a match to Jupiter's
       low-degree gravitational moments with the DFT-MD EOS.  The
       upper abscissa gives the relative mass enclosed within a
       cylinder of radius $l$.
\label{fig3}}
\end{figure}

If Jupiter formed in a region of the solar nebula where temperatures
were too low for H$_2$O to be in the gas phase, it would follow that a
primordial core would include most of the H$_2$O in solid form, and
that accreted nebular gas would accordingly be depleted in
H$_2$O. This is consistent with the last model in Table~\ref{tab1}. We
estimate that the non-H-He component of Jupiter's mantle comprises 2
$\pm$ 2 $M_{\rm E}$, and Jupiter's dense core comprises 16 $\pm$ 2
$M_{\rm E}$. The error bars include systematic and statistical
uncertainties in the DFT-MD EOS. The dense core is modeled as rock and
H$_2$O in solar proportions, but could for example include $\sim 5
M_{\rm E}$ of sedimented He for an initial $Y=0.25$.
Simulation parameters were chosen in order to reach an accuracy of 1\%
within the DFT method. A uniform 1\% alteration of the pressure
changes the core mass by 1 $M_{\rm E}$. An earlier Jupiter model based
on a completely different set of DFT-MD EOS results for pure
hydrogen~\citep{Vo07} yielded a core mass that differed by 2 $M_{\rm
E}$. We use this difference to estimate the uncertainties of our
predictions.

An acceptable model of Jupiter's interior must match the observed
constraints of the planet's mass, equatorial radius at a standard
pressure of 1 bar, and the observed multipole moments $J_2$, $J_4$,
$J_6$ of the gravity field normalized to the equatorial
radius. Jupiter's multipole moments are primarily a response to the
planet's rotation, and the higher-order moments can be strongly
affected by nonuniform rotation. Traditional models of Jupiter's
interior structure and external gravity have assumed solid-body
rotation equal to the rotation rate of the magnetic field (the
corresponding rotation period is 9:55:29.7 hours, stable over
decades~\citep{citation20}). Our model calculations find the
axially-symmetric mass density through a self-consistent field (SCF),
calculated to fourth order in the rotational
distortion~\citep{citation21,citation22}.  We incorporate differential
rotation on cylinders to the same order as the SCF.  Table~\ref{tab1}
shows that models with solid-body rotation and the DFT-MD EOS predict
a $|J_4|$ that is seven standard deviations larger than the observed
$|J_4|$. The error bar of $J_4$ has decreased by about a factor
three~\citep{citation24} with respect to earlier
determinations~\citep{citation23}. As predicted by
theory~\citep{citation25}, reduced $|J_4|$ is associated with a more
rapid decrease of density with radius in the pressure range 1 to 100
GPa. However, we have no physical justification for making ad hoc
modifications to the $P(\rho)$ relation in this pressure range. Since
we have a fully convective mantle in our model, we cannot match $J_4$
by distributing helium unevenly in an upper and lower mantle layer,
which is one major difference compared to earlier models
e.g.~\citep{citation6}. Because the mantle metallicity is so low,
neither can we redistribute mass by invoking composition gradients.
It is more plausible that Jupiter's $J_4$ is affected to a measurable
extent by zonal winds in this pressure range.  We bring the calculated
$J_4$ into agreement without appreciably changing any other parameters
of the model by assuming the zonal wind profile shown in
Fig.~\ref{fig3} (preferred model in Table~\ref{tab1}). The shallower
part of the wind profile could be adjusted further to force better
agreement with $J_6$, but such a refinement lies outside the scope of
this paper. In that regard, we note a recent paper~\citep{citation26}
that models deep zonal flows in Saturn by fitting gravity data.

To ensure that numerical errors do not play any role in the mismatch
of $J_4$, we checked our calculations with two independent theories
for rotationally-distorted models of Jupiter.  The standard
theory~\citep{citation32} solves iteratively for the shape of level
surfaces to third order in the rotational perturbation, sufficient to
calculate $J_2$ to a relative precision of 10$^{-3}$, $J_4$ to about
10$^{-2}$, and $J_6$ to about 10$^{-1}$. The results agree well with
the more accurate SCF method.

It has been shown~\citep{citation27} that high-order Jupiter gravity
harmonics are sensitive to interior dynamics in the outermost
planetary layers at pressures less than about 10 GPa.  There is no
significant uncertainty in the hydrogen EOS at these pressures, which
are accessible to experiment~\citep{citation33}.  This is why we
investigated whether the DFT-MD EOS, coupled with plausible models of
Jovian mantle dynamics, could reproduce the observed Jovian gravity
field.  According to the Poincar\'e-Wavre theorem~\citep{citation34},
the requirement that there be a single pressure-density relation
(barotrope) within Jupiter is a necessary and sufficient condition
that deep windspeeds be constant on cylindrical surfaces of constant
$l$.  Mapping the observed windspeeds onto such cylinders and then
fitting the implied centrifugal potential with a polynomial to degree
eight in $l$, we obtain the heavy curve in Fig.~\ref{fig3}. When this fit is
incorporated in the SCF calculations, we find insignificant shifts in
the predicted values of $J_n$ ($n = 2, 4, 6$), although the winds
could affect higher-order Jupiter gravity
components~\citep{citation27}.
The assumed retrograde deep zonal wind has negligible effect on
Jupiter's total spin angular momentum, reducing the latter by only
0.06\% with respect to a windless model.

New observational data concerning these theoretical predictions are
expected from the NASA mission Juno. Juno is a low-periapse Jupiter
orbiter that will return unprecedented data on Jupiter's high-order
gravitational and magnetic fields during 2016. It may present the
first direct evidence of deep interior zonal flows in Jupiter proposed
here.

%%%%%%%%%%%%%%%%%%%%%%%%%%%%%%%%%%%%%%%%%%%%%%%%%%%%%%%%%%%%%%%%%%%%

\acknowledgments

B.M., W.B.H., and J.V. acknowledge support from NASA and NSF. I.T.
and S.A.B. acknowledge support from NSERC, and computational resources
from IRM (Dalhousie) and Westgrid. We thank A. Burrows and J. Fortney
for comments.

%\bibliographystyle{apj}
%\bibliography{jupiter}

%\clearpage

\clearpage

\widetext
\begin{table}
%\tabletypesize{\scriptsize}
\caption{Constraints on Jupiter interior structure and parameters of 
           models based on DFT-MD EOS.  Predicted conditions at the core-mantle boundary (CMB) are listed last.
           \label{tab1}}

\begin{tabular}{p{115pt} | p{45pt} | p{42pt}  p{42pt}  p{42pt} | p{32pt} | p{32pt} | p{30pt} | p{30pt} }
%\tablewidth{0pt}
%\tablehead{
\hline
   Model & Equatorial radius (km) & 
   $J_2$$\times$$10^6$ $\left.\right.$~~~~~~~Gravity~moments$\!\!\!\!\!\!\!\!\!\!\!\!\!\!\!\!\!\!\!\!\!\!\!\!\!\!\!\!\!\!\!\!\!\!\!\!\!\!$ &  
   $J_4$$\times$$10^6$ &
   $J_6$$\times$$10^6$ &
   Core mass ($M_E$) & Mantle ice ($M_E$) & $T\,$(K) at CMB & $P\,$(GPa) at CMB\\[5pt]
%}
\hline
%\startdata
%\colline
Observed &	~~~71492 &	14696.43 $\pm$~0.21 &	$-$587.14 $\pm$~1.68 &	34.25 $\pm$~5.22 & $\ldots$ & $\ldots$ & $\ldots$ & $\ldots$\\[15pt]
Solid-body rotation & ~~~71493 & 14718	& $-$620 & 37.5 &	16.7 & 2.1 &	12500	& 3800\\[15pt]
Average surface~winds &	~~~71493	& 14737 & $-$623 & 37.3	& 16.7 & 2.1 & 12500 & 3870\\[15pt]
Preferred model: deep winds & ~~~71492 & 14697.14	& $-$586.3 & 23.9 & 16.2 & 2.1 &	12500 &	3800\\
\hline
%\enddata
\end{tabular}
\end{table}

\end{document}